\documentclass[12pt,a4paper]{article}

\usepackage{indentfirst}
\usepackage {amsfonts}

\def\tabaddress#1{{\small\it\begin{tabular}[t]{c}#1 \\[1.2ex]\end{tabular}}}
\def\UPCMAT{Departamento de Matem\'atica Aplicada IV\\
   Campus Norte U.P.C., Ed. C-3\\
   C/ Jordi Girona 1.
   E-08034 Barcelona, Spain}

\def\thebiblio#1{
\begin{center}\bf \large References
\end{center}
\list
{[\arabic{enumi}]}{\settowidth\labelwidth{#1.}\leftmargin\labelwidth
 \advance\leftmargin\labelsep
 \usecounter{enumi}}
 \def\newblock{\hskip .11em plus .33em minus -.07em}
 \sloppy
 \sfcode`\.=1000\relax}

\newtheorem{teor}{Theorem}

\newtheorem{corol}{Corollary}

\def\beq{\begin{equation}}
\def\eeq{\end{equation}}
\def\bea{\begin{eqnarray}}
\def\eea{\end{eqnarray}}
\def\beann{\begin{eqnarray*}}
\def\eeann{\end{eqnarray*}}
\def\beasn{\begin{sneqnarray}}
\def\eeasn{\end{sneqnarray}}
\def\ben{\begin{enumerate}}
\def\een{\end{enumerate}}
\def\bit{\begin{itemize}}
\def\eit{\end{itemize}}
\def\dst{\(\displaystyle}

\def\derpar#1#2{\frac{\partial{#1}}{\partial{#2}}}

\def\qed{\ifvmode\removelastskip\fi
{\unskip\nobreak\hfil\penalty50\hbox{}\nobreak\hfil
\hbox{\vrule height1.2ex width1.2ex}\parfillskip=0pt
\finalhyphendemerits=0 \par\smallskip}}
\font\fr=eufm10 scaled \magstep 1 %(caracteres goticos)
\font\es=msbm11                   %(caracteres ``doble barra´´)
\def\vf{\mbox{\fr X}}
\def\df{{\mit\Omega}}

\def\d{{\rm d}}

\def\Real{\mbox{\es R}}

\def\Zahl{\mbox{\es Z}}

\def\Tan{{\rm T}}
\def\Lie{{\rm L}}
\def\inn{{i}}
\def\Cinfty{{\rm C}^\infty}

\def\qed{\ifvmode\removelastskip\fi 
{\unskip\nobreak\hfil\penalty50\hbox{}\nobreak\hfil 
\hbox{\vrule height1.2ex width1.2ex}\parfillskip=0pt 
\finalhyphendemerits=0 \par\smallskip}} 
\def\dif{{\rm d}} 
\def\deriv{\@ifnextchar[{\@deriv}{\@deriv[]}} 
   \def\@deriv[#1]#2#3{\mathchoice% 
{{\dif^{#1}#2\over\dif{#3}^{#1}}}{{\dif^{#1}#2/\dif{#3}^{#1}}}% 
{{\dif^{#1}#2\over\dif{#3}^{#1}}}{{\dif^{#1}#2/\dif{#3}^{#1}}}} 
\def\derpar#1#2{\mathchoice% 
{{\partial#1\over\partial#2}}{{\partial#1/\partial#2}}% 
{{\partial#1\over\partial#2}}{{\partial#1/\partial#2}}}

\def\secteqno{\@addtoreset{equation}{section}% 
\def\theequation{\thesection.\arabic{equation}}} 
\newcounter{subequation} 
\def\thesubequation{\alph{subequation}} 
\def\sneqnarray{\stepcounter{equation}\let\@currentlabel=\theequation 
\setcounter{subequation}{1} 
\def\@eqnnum{{\rm (\theequation.\thesubequation)}} 
\global\@eqcnt\z@\tabskip\@centering\let\\=\@eqncr\let\@@eqncr=\@@sneqncr 
$$\halign to \displaywidth\bgroup\@eqnsel\hskip\@centering 
 $\displaystyle\tabskip\z@{##}$&\global\@eqcnt\@ne 
 \hskip 2\arraycolsep \hfil${##}$\hfil 
 &\global\@eqcnt\tw@ \hskip 2\arraycolsep $\displaystyle\tabskip\z@{##}$\hfil 
  \tabskip\@centering&\llap{##}\tabskip\z@\cr} 
\def\endsneqnarray{\@@sneqncr\egroup $$\global\@ignoretrue} 
\def\@@sneqncr{\let\@tempa\relax 
   \ifcase\@eqcnt \def\@tempa{& & &}\or \def\@tempa{& &} 
   \else \def\@tempa{&}\fi 
     \@tempa \if@eqnsw\@eqnnum\stepcounter{subequation}\fi 
     \global\@eqnswtrue\global\@eqcnt\z@\cr} 
\def\nobiblabels{\def\@lbibitem[##1]##2{\@bibitem{##2}}} 
\makeatother 
\def\buildord#1\over#2{\mathord{\mathop{\kern0pt #2}\limits^{#1}}} 
 
\def\Ker{\mathop{\rm Ker}\nolimits}

\def\transp#1{{}^{t}\kern-.15em\relax#1} 
\def\Tens^#1{\buildord #1\over\otimes } 
\def\inn{\mathop{\rm i}\nolimits} 
\def\Img{\mathop{\rm Im}\nolimits} 
\def\Cinfty{{\rm C}^\infty} 
\def\Tan{\mathrm{T}}

\def\bff{{\bf f}}

\def\bfu{{\bf u}}
\def\bfx{{\bf x}}
\def\bfy{{\bf y}}
\def\bfF{{\bf F}}
\title{\textbf{ON SOME ASPECTS OF THE \\ GEOMETRY OF DIFFERENTIAL \\ EQUATIONS IN PHYSICS}}
\author{\sc Xavier Gr\`acia$^{1}$ , \sc Miguel C. Mu\~noz-Lecanda$^{2}$ , \\
\sc Narciso Rom\'an-Roy$^{3}$
\\
\tabaddress{\UPCMAT}}
\begin{document}

%\pagestyle{myheadings}
%\markright{\sc X. Gr\`acia {\it et al\/},
%   \sl On some geometric aspects of differential equations.}

\maketitle

\begin{abstract}
\normalsize{
In this review paper,
we consider three kinds of systems of differential equations,
which are relevant in physics, control theory and other
applications in engineering and applied mathematics;
namely: Hamilton equations, singular differential equations,
and partial differential equations in field theories.
The geometric structures underlying these
systems are presented and commented. 
The main results concerning these structures are stated and discussed, 
as well as their influence on the study 
of the differential equations with which they are related.
Furthermore, research to be developed in these areas is also commented.
}
\end{abstract}

\vspace{1cm}

\begin{quotation}
\end{quotation}

\vspace{1cm}

\vbox{\raggedleft AMS s.\,c.\,(2000):
34A09, 34A26, 35Q99, 37J05, 53D05, 53D35, 70H05.
 \\
PACS (1999):
 02.40.Hw, 03.50.Kk, 11.10.Ef, 45.20.Jj, 45.30+s.
}\null

\vspace{1cm}
\begin{itemize}
\item[1] {\small {\bf e}-{\it mail}: xgracia@ma4.upc.es}
\item[2] {\small {\bf e}-{\it mail}: matmcml@ma4.upc.es}
\item[3] {\small {\bf e}-{\it mail}: matnrr@ma4.upc.es}
\end{itemize}

\thispagestyle{empty}

\clearpage

\tableofcontents

\section{Introduction}

The aim of this paper is to make evident the relation 
between some kinds of systems of differential equations 
and certain underlying (hidden) geometric structures, 
whose analysis may allow us to clarify the properties of the
system described by those equations.

We pay attention to three different situations.
Two of them concern ordinary differential equations ({\sc ode}s), and
the third one is partial differential equations ({\sc pde}s).

The first one is (autonomous) Hamiltonian systems,
with the  symplectic geometry as background structure.
In fact, there is a historical interplay between the equations
and the geometric structure, and results from one of these aspects depend for their
interpretation on the other, leading to new insights in both aspects.
This subject is treated in Section~\ref{hamilton}.

The second is singular differential equations;
that is, those which cannot be written in normal form;
and in particular, the case where the dependence
on the derivatives is linear.
The underlying structure is a submanifold of a tangent bundle.
The study of systems of singular differential equations was made separately
in theoretical physics and in some technical areas such as engineering
of electric networks or control theory. In both cases, algorithms for
solving the problems (which are essentially the same) were developed,
although independently. The study of the underlying geometric structure
has proved to be invaluable to find the solutions
of these equations and understand their properties.
All of this is developed in Section \ref{singular}.

Finally, the third subject presents a different situation,
and concerns partial differential equations
appearing in physics and technical applications.
Many partial differential equations admit a variational formulation.
Multisymplectic geometry underlies this formulation.
It seems to be a generalization of symplectic geometry,
but the difficulty of the problems is of increasing magnitude.
One of the objectives of this lecture is to present
some of these problems, enabling us to clarify
some aspects related to all these partial differential equations.
Section \ref{field} is devoted to it.

 Manifolds are supposed to be real, paracompact,
 and $\Cinfty$. Maps are $\Cinfty$. Sum over crossed repeated
 indices is understood.

\section{Hamilton equations and symplectic geometry}
\protect\label{hamilton}

\subsection{Hamilton equations}

The study of (time-independent) Hamiltonian systems is a classical subject
which arises from the transformation of the  Euler-Lagrange equations,
a second-order system of {\sc ode}s, into a first-order system,
using the so-called {\sl Legendre transformation}
\cite{AM}, \cite{Ar}, \cite{Wh}.

For a system defined by a  Lagrangian function
$L(t,q^i,\dot q^i)$, the  Euler-Lagrange equations are
\beq
[L]_{i} := \derpar{L}{q^i}-\frac{\d}{\d t}\left(\derpar{L}{\dot q^i}\right) = 0
\ ,\ i=1,\ldots ,m
\label{eleqs}
\eeq
which is a second-order system for $(q^1,\ldots ,q^m)$.
The associated Hamiltonian system introduces new variables $(p_1,\ldots ,p_m)$
called {\sl momentum coordinates}, and the so-called {\sl Hamiltonian function},
defined as $H(t,q^i,p_i)=\dot q^ip_i-L(t,q^i,\dot q^i)$,
where $(\dot q^1,\ldots ,\dot q^m)$ are supposed to be solved from the relations
$$
p_i=\derpar{L}{\dot q^i} \ ,\ i=1,\ldots ,m
$$
With all these data, the
 Hamilton equations (Lagrange, 1808) are
\beq
\dot q^i=\derpar{H}{p_i}\ , \ \dot p_i=-\derpar{H}{q_i}
\ ,\ i=1,\ldots ,m
\label{heqs}
\eeq
It is well known that both systems (\ref{eleqs}) and (\ref{heqs}) are
equivalent in the following sense: if $(q^1(t),\ldots ,q^m(t))$ is a
solution of (\ref{eleqs}), then $(q^1(t),\ldots ,q^m(t);p_1(t),\ldots ,p_m(t))$,
with \dst p_i(t)=\derpar{L}{\dot q^i}(t,q^i(t),\dot q^i(t))\),
is a solution of (\ref{heqs}), and conversely.
The procedure of transforming Euler-Lagrange's equations into Hamilton's equations
is known as {\sl Legendre transformation}.

This situation is the usual for classical systems.
Nevertheless, there is a large class of interesting physical models
(mainly related with relativistic systems)
for which this procedure is not possible.
They are called {\sl singular} systems, and
will be discussed in Section \ref{singular}. 

It is important to point out that there are also dynamical systems
of Hamiltonian type (i.e., whose dynamical equations are of the form
(\ref{heqs})) which have no Lagrangian counterpart; that is,
they are not defined by a Lagrangian function,
but by giving a Hamiltonian function (see, for instance, \cite{Sou}).

\subsection{Geometric formulation of Hamilton's equations}

At this point we can ask about the geometric structure underlying
these systems of {\sc ode}s.
 For simplicity we can only consider the time-independent case.
In the first case, we have the tangent bundle
$\Tan Q$ of the manifold $Q$ which represents the 
 configuration space of the system. The tangent bundle has some 
natural geometric structures. Among them,
the  vertical endomorphism and the  Liouville vector field
can be used to construct several geometric objects that allow us to express
the Euler-Lagrange equations in an intrinsic form.
 Nevertheless, it seems that the geometric structure underlying the
Hamilton equations is more interesting for many applications.
In order to obtain it, observe that these equations
can be written in matrix form
$$
(\dot q^i,\dot p_i)\left(\matrix{ 0 & {\rm I} \cr -{\rm I} & 0 \cr}\right)=
\left(\derpar{H}{q^i},\derpar{H}{p_i}\right) \
$$
The skew-symmetric matrix 
\dst\omega=\left(\matrix{ 0 & {\rm I} \cr -{\rm I} & 0 \cr}\right)\)
can be interpreted as a $2$-form
$\omega=\d q^i\wedge\d p_i$,
and the equation means that the contraction of $\omega$ with the velocity vector
\dst\dot q^i\derpar{}{q^i}+\dot p_i\derpar{}{p_i}\)
equals the $1$-form \dst\d H=\derpar{H}{q^i}\d q^i+\derpar{H}{p_i}\d p_i\).
In intrinsic terms, $(q^i(t),p_i(t))$ is the local representation
of a path $\xi$ in a manifold $M$ (the  phase space of the system),
endowed with a $2$-form
$\omega$, and $\xi$ satisfies the differential equation
$$
\inn_{\dot\xi}\omega=\d H\circ\xi
$$
where $H\colon M\to\Real$ is a function.
Furthermore, $\omega$ is required to be symplectic; that is, closed and
non-degenerate. So we reach the concept of {\sl symplectic manifold}
$(M,\omega)$, which is a geometric structure similar to 
that of a Riemannian manifold.
In a symplectic manifold, given a function $f\in\Cinfty(M)$, there exists a unique
vector field $X_f\in\vf(M)$ associated with $f$, such that $\inn_{X_f}\omega=\d f$.
$X_f$ is called the {\sl Hamiltonian vector field} associated with $f$, and
$f$ is called the {\sl Hamiltonian function} associated with $X_f$.
Hence, note that the Hamilton equations for $\xi$ can be written as
$\dot\xi=X_H\circ\xi$, where $X_H$ is the Hamiltonian vector field for $H$.

Elementary examples of symplectic manifolds are
$\Real^2$ and $S^2$, both with the area element $\omega$,
and the  cotangent bundle
$\Tan^*Q$ of the configuration manifold $Q$ of the system,
which carries a canonical symplectic structure.
In this last case, the above mentioned Legendre transformation is a map
${\cal F}L\colon\Tan Q\to\Tan^*Q$.

As we will see in Section \ref{singular}, this geometric framework
can be extended to describe singular Hamiltonian systems,
just allowing the closed form $\omega$ to be degenerate
(then it is called a presymplectic form,
and $(M,\omega)$ is said to be a {\sl presymplectic manifold}).

An skew-symmetric bilinear map on $\Cinfty(M)$, which is called the {\sl Poisson bracket},
can be defined in every symplectic manifold $(M,\omega)$ as follows
$$
\{ f,g\}:=\omega(X_g,X_f)
$$
which allows us to write the Hamilton equations in the form
$$
\dot q^i=\{ q^i,H\} \ , \ \dot p_i=\{ p_i,H\}
$$
(Actually, the concept of Poisson bracket is more general than the symplectic structure
\cite{LM-87}, \cite{Ma-92}).

A relevant feature of Hamiltonian vector fields is that they preserve
the symplectic (resp. presymplectic) form. 
In fact, $X\in\vf(M)$ is a (locally) Hamiltonian vector field
iff $\Lie_X\omega=0$. In the simple example of the symplectic manifold $(\Real^2,\omega)$,
this result means that the area of
a region $U\subset\Real^2$ and the area of the transformed region $\varphi(U)\subset\Real^2$,
under the flow of a Hamiltonian vector field $X$, are the same.
This flow is an example of a map preserving the symplectic form,
that is, a  symplectomorphism.

\subsection{Main results concerning symplectic geometry}

In 1912, Poincar\'e,
working on the symplectic manifold $(\Real^2,\omega)$,
proposed a simplified $2$-dimensional model for the solar system,
obtaining the first result on symplectic geometry which was used for studying
Hamiltonian systems (see \cite{Ar}, \cite{Can}):

\begin{teor}
{\rm (Poincar\'e's last geometric theorem)}.
Suppose $\varphi\colon A\to A$ is an area-preserving
diffeomorphism of the closed annulus $A=\Real/\Zahl\times[-1,1]$,
which preserves the two components of the boundary, and twists them in opposite
directions. Then $\varphi$ has at least two fixed points.
\end{teor}

This result leads to the concept of Poincar\'e map for Hamiltonian
systems, and to determining the existence of periodic orbits for
many dynamical systems. There is a generalization of this result
(the so-called {\sl Arnold's conjecture})
and several applications of it (\cite{Can}, pp. 55-56).
In particular, the relation between the number of fixed points
of a symplectomorphism and the critical points of a
{\sl Morse function} in the manifold \cite{Ar}, \cite{Sa}.

This was the standpoint of a long sequence of results on
the geometry and topology of symplectic manifolds,
whose developments have led to a clarification of the
structure and properties of Hamiltonian systems.
Some of them are the following
(see, for instance, \cite{AM}, \cite{LM-87}, \cite{Sa}, \cite{We} for details):

\begin{teor}
{\rm (Darboux, 1882)}.
Let $(M,\omega)$ a symplectic manifold, and $x\in M$. Then, there exists
a local chart $(U;q^i,p_i)$ at $x$ such that 
$\omega\vert_U=\d q^i\wedge\d p_i$.
\end{teor}

Some consequences of this theorem (which, in fact, predates
the Poincar\'e theorem) are:
\ben
\item
The Hamiltonian systems are locally equivalent in the following sense:
given two Hamiltonian vector fields, there is a symplectic transformation
that maps locally one into the other.
\item
There are no local invariants in symplectic geometry
(such as curvature in Riemannian geometry).
\item
The group of diffeomorphisms preserving the symplectic structure is infinite dimensional.
\een
An analogous theorem and similar consequences can be stated
for presymplectic manifolds. 

\begin{teor}
{\rm (Marsden-Weinstein's reduction theorem, 1974. \cite{MW-74})}.
Let $G$ be a Lie group which acts symplectomorphically in a
symplectic manifold $(M,\omega)$, with associated
{\rm momentum mapping} $J\colon M\to {\bf g}^*$
(where ${\bf g}^*$ denotes the dual of the Lie algebra of $G$). As $J$
is equivariant, if $0\in{\bf g}^*$ is a regular
value of $J$, then $J^{-1}(0)$ is a submanifold of $M$
which is invariant under the action of $G$.
Furthermore, if the action of $G$ is free and proper,
then $J^{-1}(0)/G$ is a symplectic manifold with dimension equal to
$\dim\, M-2\dim\, G$.
\end{teor}

The origin of this reduction procedure is very old.
The original ideas come from Euler, Lagrange, Jacobi, Poisson, Lie and Noether.
In modern formulation, Moser, Arnold, Guillemin, Sternberg, Marle
and many others have contributed to its development and applications
(see \cite{MW} and references therein).

Thus, reduction theory concerns the removal of variables
using symmetries and conservation laws and,
as a consequence of the theorem, every Hamiltonian system with symmetries
can be reduced to another Hamiltonian system.
In addition, other relevant consequences are the
{\sl Atiyah-Guillemin-Sternberg} and {\sl Delzaut theorems}
\cite{Can}.

As above, this theorem can be generalized for presymplectic manifolds,
as well as to other different singular cases
(see, for instance, \cite{EMR-99} and the references quoted therein). 

\begin{teor}
{\rm (Kostant-Kirillov-Souriau, 1970. \cite{Kos}, \cite{Sou})}.
Let $G$ be a Lie group, and ${\cal G}$ its Lie algebra.
The orbits of the coadjoint action of $G$ in ${\cal G}^*$
are endowed with a canonical symplectic structure
$\omega^*\in\df^2({\cal G}^*)$ which is $G$-invariant.
\label{KKS}
\end{teor}

This theorem has some important corollaries:

\begin{corol}
If the action of $G$ on $(M,\omega)$ is Poissonian
(see \cite{LM-87}), then:
\ben
\item
The associated momentum mapping $J$ maps the orbits
${\cal O}_x$ of the action of $G$ in $M$ into the orbits
${\cal O}^*_{J(x)}$ of the coadjoint action of $G$ in ${\cal G}^*$.
Furthermore, $J$ maps the symplectic form $\omega$ into the natural symplectic
form of ${\cal O}^*_{J(x)}$.
\item
If the action is also transitive, then $J$ is a local symplectomorphism.
\een
\end{corol}

Roughly speaking, this means that Hamiltonian systems with symmetries
are the orbits of the coadjoint action.

Another subject related to this theorem is the so-called
{\sl geometric quantization}: in a natural way,
a Hilbert space ${\cal H}$ can be associated with every symplectic manifold.
Therefore, the phase space and the state variable functions are transformed into
a Hilbert space and self-adjoint operators on it
(see, for instance, \cite{EMR-gq}, \cite{Ki-gq}, \cite{Ko-70},
\cite{Sn-80}, \cite{So-66}, \cite{Wo}).

\begin{teor}
{\rm (Arnold-Liouville, 1969. \cite{Ar})}.
Let $(M,\omega,H)$ be a $2n$-dimensional integrable system with
integrals of motion $f_1=H,f_2,\ldots ,f_n$. Let $c\in\Real^n$ be
a regular value of $f=(f_1,\ldots,f_n)$.
The corresponding level set $f^{-1}(c)$ is a Lagrangian submanifold
of $M$.
\ben
\item
If the flows of $X_{f_1},\ldots,X_{f_n}$ starting at a point $p\in f^{-1}(c)$
are complete, then the connected component of $f^{-1}(c)$ containing
$p$ is a homogeneous space for $\Real^n$. With respect to this affine
structure, the component has coordinates $\varphi_1,\ldots,\varphi_n$,
known as {\rm angle coordinates}, in which the flows of the vector fields
$X_{f_1},\ldots,X_{f_n}$ are linear.
\item
There are coordinates $\psi_1,\ldots,\psi_n$, which are known as
{\rm action coordinates}, complementary to the {\rm angle coordinates},
such that the $\psi_i$'s are integrals of motion and 
$(\varphi_1,\ldots,\varphi_n,\psi_1,\ldots,\psi_n)$ form a Darboux chart.
\een
\label{AL}
\end{teor}

This shows that the dynamics of
integrable systems (see \cite{Can}) are very simple, and they have explicit solutions in
these  action-angle coordinates (see \cite{Can}, p.110, and \cite{CB}).

Another interesting result is the classification of 
normal forms of quadratic Hamiltonians (see \cite{Ar}, p.486),
which is based on {\sl Williamson's theorem} \cite{Wi1}: 
a result about the classification of quadratic forms in
symplectic vector spaces.

We also refer to a theorem of Banyaga which states that
some classical structures and, in particular, the symplectic form,
are determined by their automorphism groups,
the group of symplectomorphisms \cite{Ba97}.
Closely related to this result is the
{\sl Lee Hwa Chung theorem} which establishes that,
in a given symplectic manifold, appart from the symplectic
and volume forms, the only differential forms invariant
by the set of locally Hamiltonian vector fields are
multiples of exterior powers of the symplectic form.
The original version of this theorem \cite{Hw-47} concerns
the uniqueness of invariant integral forms
(the {\sl Poincar\'e-Cartan integral invariant}\/)
under canonical transformations (that is, those
diffeomorphisms in a symplectic manifold,
mapping Hamiltonian vector fields into themselves),
and this result leads to characterize canonical transformations in the 
Hamiltonian formalism of Mechanics,
identifying them with the symplectomorphisms. Afterwards,
these results were generalized to presymplectic Hamiltonian systems \cite{GLR-84}.

There are many other topics concerning symplectic geometry
in the realm of Hamiltonian systems.
For instance, the study of Lagrangian submanifolds and foliations,
which has application to several kind of problems \cite{We}; such as,
to give a very nice interpretation of the Lagrangian and Hamiltonian dynamics
and the Legendre transformation \cite{Tu-76a}, \cite{Tu-76b}, \cite{Tu-77},
to characterize the generating functions
of canonical transformations (symplectomorphisms) \cite{AM}, \cite{We}, 
or to get a geometrical framework for the Hamilton-Jacobi theory \cite{VK}.
Apart from this, we can point out the so-called
{\sl symplectic integrators}, which allow us to obtain
algorithms for the discretization of the Hamilton equations and numerical solutions
of them, and are based on the conservation of the symplectic form
\cite{BS}, \cite{CM}, \cite{LMd}, \cite{MPS}.
There is also the study of the Schr\"odinger equation 
as an infinite dimensional Hamiltonian system
\cite{Tuy} (which is related to geometric quantization).

Finally, as presymplectic geometry is the arena for singular Hamiltonian
systems, we must also mention some remarkable results concerning this topic.
One of the most important is the {\sl Weinstein extension theorem}
\cite{We-2}, and one of its consequences, the
{\sl coisotropic imbedding theorem} \cite{Go-82}, \cite{Ma-83},
which allow to establish the local structure of presymplectic Hamiltonian systems
as systems defined in a symplectic manifold \cite{CGIR-85}, \cite{Sni-74}.

\section{Singular differential equations}
\protect\label{singular}

\subsection{Systems described by singular differential equations}

In general, an ordinary differential equation is a relation 
$\bfF(t,\bfx,\dot\bfx,\ddot\bfx,\ldots)=0$ 
involving an independent variable~$t$, 
a dependent variable $\bfx(t)$, 
and its derivatives up to a certain order.  
When the highest-order derivative can be solved, say 
$\bfx^{(k)} = \bff(t,\bfx,\ldots)$, 
the equation is said to be in normal form, 
which is the most appropriate for studying and solving the equation, 
especially thanks to the  existence and uniqueness theorem.
However, we are mainly interested in the case 
where this highest-order derivative cannot be solved, 
not only on a point, but on an open set, 
and this is what we will mean with the term ``singular''.

Although of leeser importance than equations in normal form, 
singular differential equations have been studied for decades, 
especially in the last 30 years. 
The main reason is that they appear in many applications, 
under various names: 
singular, degenerate or implicit systems, 
differential-algebraic equations ({\sc DAE}s), 
descriptor systems, \ldots\ 
Let us consider some of these applications. 
\begin{itemize}
\item
{\sl Theoretical physics}. 
A Lagrangian $L(t,q^i,\dot q^i)$ is called {\sl singular} 
when its hessian matrix 
\dst\frac{\partial^2L}{\partial\dot q^i\,\partial\dot q^j}\) 
is singular. 
This means that its Euler--Lagrange equations (\ref{eleqs})
are singular. 
Such Lagrangians appear in the description of relativistic phenomena
and, as a consequence, in all the fundamental theories of physics; 
they were first studied by Dirac and Bergman in 1950.
(See, for instance, \cite{Ca-sl}). 
\item 
{\sl Applied mathematics}. 
In control theory, it is usual to deal with
linear singular systems; as for instance 
$$ 
E \dot \bfx = A\bfx+ B\bfu , 
\qquad 
\bfy = C\bfx + D\bfu , 
$$ 
where $E$, $A$, $B$, $C$ and $D$ are 
constant matrices, $E$ singular.  
Such systems appear in various problems coming from 
optimal control, constrained control and electric circuits, 
and so they have been widely studied. 
One of the first books on these systems was
\cite{Ca-80}; more recently \cite{Kac} 
gives a complete account of singular linear systems. 
The nonlinear case has been less studied, 
and is related to the theory of singular perturbations 
\cite{Isi}.
\item
{\sl Other applications}. 
Many of them can be found in circuit theory, 
but also in chemical and industrial engineering.
We also know of applications to more faraway fields, 
such as econometry or biology.
\end{itemize}

It is known that the introduction of additional variables 
allows us to transform any differential equation 
into a first-order autonomous equation, 
$\bfF(\bfx,\dot\bfx)=0$, 
or $\dot\bfx = \bff(\bfx)$ in normal form;.
For the sake of simplicity, 
from now on we will consider only such equations.

\subsection{Problems arising when solving singular systems}

In general, singular systems cannot be represented by a vector field,
so their integration ---analytical or numerical---
leads to new problems, 
which are not present in the regular case.
\paragraph{Consistency}
The first point to be noted is that a singular differential equation 
may not have solutions passing through 
each point of the phase space $M$ of the system.
In other words, not all values of the variables 
are admissible initial conditions.

To solve the equation, 
first one should identify these admissible initial conditions.
Usually they are not obtained directly but in an algorithmic way.
For instance, the differential equation
may imply some relations $\phi^a(\bfx) = 0$
among the variables~$\bfx$;
these relations,
sometimes called {\sl primary constraints},
define a subset $M_1 \subset M$,
the {\sl primary constraint subset}.
Hence, the first step consists in finding
these constraints.

Nevertheless, the problem does not finish at this point: 
not only the initial conditions $\bfx_\circ$ must be inside~$M_1$, 
but also their evolution $\bfx(t)$ must remain there 
---this is a tangency condition. 
This implies additional constraints, as will be shown later. 
This procedure is repeated until a
 final constraint subset $M_f$ is obtained. 
\paragraph{Uniqueness}
Under some favourable conditions of regularity, 
the preceding procedure ends with a submanifold,
where the differential equation 
can be represented by a family of tangent vector fields, 
so that their integral curves describe the solutions.
Since different vector fields yield different solutions, 
in general, giving $\bfx(0) = \bfx_\circ$,
does not determine the solution.
Sometimes, even the knowledge of all the derivatives at $t=0$ 
does not determine the solution.

This is not always the case. 
Sometimes the singularity of the initial differential equation 
arises from an``inappropriate'' choice of the variables 
(that is,the initial state space is too large to
describe the real degrees of freedom of the problem, due to the use 
of redundant coordinates); 
in such a case, the system has a unique solution
on the final constraint submanifold.
\paragraph{Reduction} 
However, in some physical problems 
the singularity of the system is a consequence of the
existence of a certain kind of internal symmetries
(called {\sl gauge symmetry}\/ in theoretical physics), 
accounting for the fact that there are different solutions 
representing the same  physical state. 
In this case, 
the differential equation is  undetermined: 
for everypoint in the final constraint submanifold which is
taken as an initial condition, the evolution of
the system is not determined because a
multiplicity of integral curves (of different
vector fields solution) pass through it.

A  reduction procedure can be used to remove this ambiguity:
different points of $M_f$ that can be reached
through different solutions beginning
at the same initial condition must be identified.
This quotient is the  reduced phase space of the system.
Then 
the physical states of the systems are identified with 
the points of the reduced phase space.
\paragraph{Control systems}
In addition to these problems, 
other questions are also interesting
when dealing with singular control systems: 
controllability, observability, reachability, 
stabilizability, realizability, optimality \ldots\ 
All these concepts,
which are well established for regular control systems,
must be reconsidered in the singular case. 
\paragraph{Numerical methods}
Here the matter is 
the design and convergence of numerical methods 
to solve implicit differential equations, 
and also their relation with singular perturbation problems. 
There are many articles and some books devoted to these problems 
(see, for instance, \cite{CHYZ}, \cite{HLR-89}, \cite{HW-91}).

\subsection{Geometric formulations of singular differential equations}

In geometric terms, 
a differential equation written in normal form is defined by a 
vector field $X$ on a manifold~$M$. 
Then the equation for a path $\gamma \colon I \to M$ 
reads
$\dot\gamma = X \circ \gamma$. 

In the same way, 
an implicit differential equation can be geometrically described by 
a submanifold $D \subset \Tan M$ of the tangent bundle of~$M$, 
and the differential equation is then expressed by the inclusion 
$$
\dot\gamma(I) \subset D .
$$ 
Even if the equation is initially set in euclidian space, 
one is usually led to work on submanifolds of the initial space, 
so the geometric framework is not only nice, but also necessary. 

Singular differential equations,
from this most general point of view,
have often been studied in the literature.
For instance, in
\cite{MT-78} and \cite{MMT-97} 
this general framework is applied to the Hamiltonian dynamics of
singular Lagrangian systems.
The article
\cite{MMT-92}
studies symmetries and constants of motion for these singular
equations,
whereas their integrability is studied in
\cite{MMT-95}. 
In the same way,
but taking $M$ as a euclidean space,
\cite{RR-94}
studies the existence of solutions of the same problem,
and also gives an algorithm for finding these solutions
under some regularity conditions; 
this article is indeed a geometric formulation of the authors'
previous works 
\cite{RR-91}, \cite{Rhe-84}.
More or less, the same algorithm is given in 
\cite{Rei}.

In the literature,
the most widespread singular differential equations
are of a special type:
they combine some restrictions on the base~$M$
and some linear relations among the velocities
in $\Tan M$.
In general, these equations can be written, in coordinates, as 
$
A(\bfx) \dot \bfx = b(\bfx) 
$,
where $A(\bfx)$ is a matrix, usually singular.
Such equations can be called {\sl linearly singular}
(they are also called {\sl quasilinear}\/). 

The geometric study of these equations has been developed
independently in the areas of Theoretical Physics and
Applied Mathematics.

In Theoretical Physics the initial problem
was to obtain a Hamiltonian
description for singular Lagrangians
\cite{Dir-lectures}. 
The geometrization of this problem was performed later 
\cite{Sni-74}, \cite{Lic-75}, \cite{MT-78}, 
and a more general framework is that of presymplectic systems
\cite{GNH-78}, 
where the equation of motion has the form
$$ 
i_{\dot\gamma} \omega = \alpha \circ \gamma, 
$$ 
where $\omega$ is a presymplectic form on a manifold~$M$ 
and $\alpha$ is a 1-form.
Hamiltonian dynamics corresponds to the case where 
$M \subset \Tan^* Q$ is a submanifold, 
$\omega$ is the pullback of 
the canonical symplectic form of~$\Tan^*Q$ to~$M$, 
and $\alpha = \dif H$, where $H$ is a Hamiltonian function. 
An algorithm was presented in the aforementioned paper
to study the existence of solutions of such a system.
Note that, since the transformation $X \mapsto i_X \omega$
is not an isomorphism,
there is no guarantee of either existence or uniqueness
of the solutions.
The same ideas can be applied to search the solutions in
the Lagrangian formalism for the singular case \cite{GN-80a};
but then the presymplectic equation must be supplemented with
a second-order condition \cite{GN-80}, \cite{MR-92}.
Another way to treat this problem is working in the manifold
$M = \Tan^*Q \times_Q \Tan Q$ \cite{SR-83}.

In a more general way, linearly singular differential equations
can be described as follows \cite{GP-92}:
a  linearly singular system is given by a
a vector bundle morphism $A \colon \Tan M \to F$ 
and a section $b$ of the vector bundle $F \to M$.
Then the differential equation reads
$$
A \circ \dot\gamma = b \circ \gamma .
$$
This problem can also be formulated in terms of vector fields.
In this article, Dirac's algorithm for presymplectic systems
\cite{GNH-78} 
is generalized to linearly singular systems.
This framework includes
applications to other problems like
singular Lagrangian formalism
and higher order Lagrangians
\cite{GPR-91}
and implicit Hamiltonian systems 
\cite{BS-01}.
Symmetries of linearly singular systems have been studied recently
\cite{GP-02}. 

A less general framework, but still including the presymplectic systems,
is presented in
\cite{MR-99},
which corresponds to a linearly singular system with $F=\Tan^* M$;
the constraint algorithm and symmetries are studied therein.

Furthermore,
some problems in circuit theory led several researchers to consider
singular differential equations.

For instance, in
\cite{Tak-76}
a {\sl constrained differential equation}\/ is a set of data
including a linear restriction on the velocities
as well as some constraints on the base manifold
obtained from a certain potential function. 
The discontinuous solutions of this problem, 
as well as its singularities, are analyzed.

The same geometric framework for constrained equations
is presented in
\cite{HB-84},
with some simplifications.
The problem is to find a curve $c\colon I \to E$ such that
$$
\Tan p \circ \dot c = \chi \circ c
\qquad
c(I) \subset \Sigma ,
$$
where $p\colon E \to B$ is a smooth map,
$\chi$ a vector field along~$p$,
and $\Sigma$ is a submanifold of~$E$ with the same dimension as~$M$.
Notice that this problem is also linearly singular.

We can also mention the paper
\cite{CO-88},
which considers a  generalized vector field, namely,
a vector bundle endomorphism $A\colon \Tan M \to \Tan M$
together with a vector field $v$ in~$M$.
The equation of motion is then
$$
A \circ \dot\gamma = v \circ \gamma .
$$
The aim of the paper is to classify the normal forms for
generalized vector fields.
Later papers have studied normal forms
\cite{Med} 
and stability 
\cite{RZ} 
of such equations.

Finally, a brief review on singular system of differential equations is presented
in \cite{GMR-96}, where their geometric features, including a geometric method
for obtaining a non-numerical solution, are analized.

\subsection{Geometric solution of singular equations}

 From a geometric viewpoint, 
the search for the solutions of an implicit equation 
given by a submanifold $D \subset \Tan M$ 
can be sketched very clearly: 
since a solution $\gamma$ satisfies $\dot\gamma(t) \in D$, 
necessarily $\gamma(t)$ belongs to 
$M_1 = \tau(D)$, 
where $\tau \colon \Tan M \to M$ is the natural projection. 
If this is a submanifold 
(the  primary constraint submanifold), 
we obtain a new implicit equation, defined by the subset  
$D_1 \subset \Tan M_1$, 
where $D_1 = D \cap \Tan M_1$. 
This is the first step of an algorithm that, 
assuming appropriate regularity conditions, 
may lead to a consistent differential equation 
on a certain submanifold~$M_{\rm f}$. 

The same algorithm applies to a linearly singular system
$(A \colon \Tan M \to F,b)$, 
where we want to solve the implicit equation 
$A \circ \dot\gamma = b \circ \gamma$. 
In this case, the equation
$A_x \cdot X(x) = \sigma(x)$ for the unknown
vector $X(x)$ can be solved only at the points 
$x\in M$ where this linear equation is consistent:
$b(x) \in \Img A_x$. 
These points constitute the set~$M_1$. 
Of course, the solutions must be contained in~$M_1$, 
so it is clear that 
they are also solutions of the linearly singular system 
$(A_1 \colon \Tan M_1 \to F_1,b_1)$ 
obtained by restricting $F$, $A$ and~$b$ to~$M_1$. 
This is essentially the first step of the constraint algorithm 
in Dirac's theory. 
Again, under appropriate regularity conditions, 
this algorithm may lead to 
a consistent linearly differential system 
on a  final constraint submanifold~$M_{\rm f}$, 
where $b_{\rm f}(x) \in \Img A_{{\rm f}x}$. 
Then the solutions of the differential equation 
correspond to the integral curves of a family of vector fields 
$X_{\rm f} + \Ker A_{\rm f}$. 
Finally, let us remark that 
this algorithm may be presented explicitly in terms of the constraints 
and vector fields in a suitable way for computation.

\section{Field theory and multisymplectic geometry}
\protect\label{field}

\subsection{Geometric formulation of field theory}

As one may see in Table \ref{table}, many of the main {\sc pde}s
 in Physics can be obtained from a Lagrangian function.
There are other interesting {\sc pde}s describing dynamical processes,
which are also variational with other conditions:
constrained problems, higher-order Lagrangians, etc.
In any case, in all of them, the underlying geometric problems 
have the same level of difficulty as those we are going to analyze next.

\begin{table}
\noindent
\begin{tabular}{|c|c|c|}
  \hline
  {\bf Name} & {\bf Equation} & {\bf Lagrangian} \\
  \hline
  Dirac & 
\dst \gamma^i\derpar{\psi}{x^i}=m{\rm I}\psi\) &
\dst  \bar\psi\gamma^i\derpar{\psi}{x^i}-\bar\psi\psi\)  \\
 & $\gamma^i$: Dirac matrices  & \\
 & I: identity matrix & \\
\hline
  Einstein & \dst{\cal R}_{\mu\nu}(g)=0\)  &
\dst R(g)\sqrt{-\det\, g}\) \\
(in vacuum) & $g$: metric ; ${\cal R}$: Ricci tensor & $R$: curvature \\
\hline
  Elasticity & 
\dst \frac{\partial^2u^\alpha}{\partial t^2}=\derpar{}{x^i}\left(\derpar{W}{u_i^\alpha}\right)\) &
 \dst\frac{1}{2}|u_t|^2-W\)  \\
& $W({\bf x},\nabla u)$: stored energy & \\
\hline
  Fluid dynamics &
 \dst\derpar{\bf u}{t}+\nabla_{\bf u}{\bf u}=-{\rm grad}\, p\) &
 \dst  \frac{1}{2}\langle{\bf u},{\bf u}\rangle\) \\
 & \dst  {\rm div}\,{\bf u}=0\) & \\
 & ${\bf u}$: fluid velocity; $p$: pressure & \\
\hline
  Klein-Gordon & 
 \dst \frac{\partial^2\psi}{\partial t^2}=\nabla^2\psi-m^2\psi\) &
 \dst  
\frac{1}{2}\left[\left(\derpar{\psi}{t}\right)^2-(\nabla\psi)^2-m^2\psi^2\right]\) \\
\hline
 Korteweg-de Vries & 
 \dst \derpar{u}{t}-6u\derpar{u}{x}+\frac{\partial^3u}{\partial x^3}=0\) & 
\dst\frac{1}{2}\derpar{\phi}{x}\derpar{\phi}{t}-\left(\derpar{\phi}{x}\right)^3-
\frac{1}{2}\frac{\partial^2\phi}{\partial x^2}\) \\
 & & with \dst u=\derpar{\phi}{x}\) \\
\hline
  Laplace & 
\dst \nabla^2\psi =0\) &
\dst \frac{1}{2}|\nabla\psi|^2\) \\
\hline
  Maxwell &
 \dst {\rm div} {\bf E}=\rho \ , \ {\rm curl}\,{\bf H}= \derpar{\bf E}{t}+{\bf j}\) &
 \dst \frac{E^2-B^2}{8\pi}-\rho\phi+{\bf j}\cdot{\bf A}\) \\
 & \dst {\rm div}\,{\bf H}=0\ ,\ {\rm curl}\,{\bf E}= -\derpar{\bf E}{t}\) & {\rm with} \\
 & ${\bf E},{\bf H}$: electric, magnetic fields &
 \dst{\bf E}=-{\rm grad}\phi-\derpar{\bf A}{t}\) \\
 & $\rho$: electric charge  &
 ${\bf B}={\rm curl}{\bf A}$ \\
 &  ${\bf j}$: electric current &
$\phi,{\bf A}$: electric, magnetic potentials
\\
\hline
  Schr\"odinger & 
\dst -\frac{1}{2m}\nabla^2\psi+V\psi=i\derpar{\psi}{t}\) &
\dst \sqrt{-1}\bar\psi\psi-\frac{1}{2m}g^{ij}\derpar{\bar\psi}{x^i}
\derpar{\psi}{x^j}-V(x^i,t)\bar\psi\psi\) \\
 & & $g$: metric\\
\hline
  Sine-Gordon & 
\dst \frac{\partial^2\psi}{\partial x\partial t}=\sin\,\psi\) &
 \dst  
\frac{1}{2}\derpar{\psi}{x}\derpar{\psi}{t}-\cos\,\psi\)  \\
  \hline
  Wave & 
\dst \frac{\partial^2\phi}{\partial t^2}=\nabla^2\phi\) &
 \dst  \frac{1}{2}(\dot\phi)^2-\|\nabla\phi\|^2\) \\
\hline
\end{tabular}
\caption{Some of the main {\sc pde}s in physics,
with their corresponding Lagrangian.}
\label{table}
\end{table}

The geometric way of describing these problems,
for first-order theories
(see, for instance, \cite{EMR-96}, \cite{Gc-73}, \cite{GMS-97}, \cite{GS-73}, \cite{KT-79},
\cite{Sa-89}), consists in considering a fibered manifold $\pi\colon Y\to X$
(where $X$ is an $m$-dimensional oriented manifold,
with volume form $\omega\in\df^m(X)$),
and the manifold of first-order jets of sections,
$J^1Y\stackrel{\pi^1}{\to}Y\stackrel{\pi}{\to}X$.
In this situation, there are natural geometric structures, such as:
the  vertical subbundle, the  vertical endomorphisms,
the  canonical structure form, the  module of total derivations,~\ldots. 
Using some of them, we can associate to
a Lagrangian function $L\colon J^1Y\to\Real$ a closed form
$\Omega_L\in\df^{m+1}(J^1Y)$ such that a section
$\psi\colon X\to Y$ is {\sl critical} for the variational problem
\dst S[\psi]=\int_M (j^1\psi)^*(L\omega)\) if, and only if,
$$
(j^1\psi)^*\inn_Z\Omega_L=0 \quad , \quad
\mbox{\rm for every $Z\in\vf(J^1Y)$}
$$
This equation 
can be written in other equivalent ways, using  $m$-vector fields
or  Ehresmann connections \cite{EMR-98}, \cite{LMM-96}.
In a local chart of adapted natural coordinates in $J^1Y$, $(x^\alpha, y^A,v^A_\alpha)$, the
expression of this condition is the well known system of  Euler-Lagrange equations
for the multiple variational case:
$$
 \derpar{L}{y^A}\Big\vert_{j^1\psi}-
\derpar{}{x^\alpha}\left(\derpar{L}{v_\alpha^A}\Big\vert_{j^1\psi}\right) = 0
 \quad ; \quad A=1,\ldots ,N
 $$
It is interesting to point out that, in the particular case with
${\rm dim}\, X=1$, this framework describes the  time-dependent mechanics
\cite{LMMMR}, \cite{MS-98}, and then the multisymplectic form
reduces to be a {\sl contact} or a {\sl cosymplectic form}.

As in the case of {\sc ode}s, a Hamiltonian formulation is also possible
in this case \cite{CCI-91}, \cite{EMR-00}, \cite{HK-02},
\cite{MS-99}, \cite{PR-02}, \cite{Sd-95},
after defining the corresponding  Legendre map.
In all of them, the so-called {\sl multimomentum phase space}
is a fiber bundle over $Y$ endowed with a
multisymplectic structure, which is canonical in some cases,
although in others it is constructed using additional elements
(connections or sections in the bundle). The {\sc pde}s obtained
in these formalisms are called the
{\sl Hamilton-De Donder-Weyl equations}.

To sum up, in all these formulations the underlying geometric structure
is a couple $(M,\Omega)$, where $M$ is a differentiable manifold and
$\Omega$ is a closed $k$-form ($k>2$),
and is called a {\sl multisymplectic manifold}
if $\Omega$ is $1$-nondegenerate.

Natural examples of this structure are $\Real^n$ and $S^n$ ($n>2$)
with their volume elements. Canonical models of multisymplectic
manifolds are a generalization of the
cotangent bundle of a manifold $Q$: the  multicotangent bundle
$\Lambda^k\Tan^*Q$ ($1<k\leq\dim\,Q$); that is,
the bundle of exterior $k$-forms on $Q$, and the subbundles of it
(bundles of forms). All of them are endowed with a
natural multisymplectic $(k+1)$-form (i.e.; closed and $1$-nondegenerate).
Other examples are  semisimple Lie groups,  cosymplectic manifolds
and  Calabi-Yau manifolds (see \cite{Ib-2000}).

Finally, it must be remarked that multisymplectic geometry is
not in fact the only geometrical framework
for describing these kinds of variational systems of {\sc pde}s,
and alternative, sometimes equivalent, geometric structures are
possible: the so-called {\sl polysymplectic} \cite{Gu-87}, \cite{Ka-98},
{\sl $k$-symplectic} \cite{Aw-92}, \cite{No-93}, {\sl $k$-cosymplectic} \cite{LMS-99},
and {\sl Lepagean} \cite{Kr-02} formalisms.

\subsection{Results and open problems on multisymplectic geometry}

Starting from a multisymplectic manifold $(M,\Omega)$, one can expect to
obtain results concerning the geometric structure of $M$ and the {\sc pde}s defined on $M$,
in an analogous way, as in the case of symplectic geometry.
Research on these topics has just started (see, for instance,
\cite{CIL-96}, \cite{CIL-98}, \cite{Ib-2000}), and
many of those problems are still unsolved. Next we review some of these
problems and their current status.

The first fundamental result in symplectic geometry was the Darboux theorem,
which stated that all symplectic manifolds are locally isomorphic.
This result also holds in some particular cases of multisymplectic geometry;
for instance, when the degree of $\Omega$ is equal to $\dim\, M$ (volume forms).
Nevertheless, in the general case, a multisymplectic manifold
does not admit a system of Darboux coordinates for the multisymplectic form \cite{Ma70}.
In fact this is a problem arising from linear algebra: the classification
of skew-symmetric tensors of degree greater than two is still an open problem.
The kind of multisymplectic manifolds
admitting Darboux coordinates has been identified recently \cite{CIL-98}, \cite{LMS-2003},
and they are those being locally multisymplectomorphic to bundles of forms.

%
%It is just for this class of multisymplectic manifolds that
%a partial generalization of Lee Hwa Chung theorem has been proved
%and used to characterize the group of multisymplectic diffeomorphisms 
%(which acts transitively on the manifold), 
%showing that the graded Lie algebra of infinitesimal automorphisms
%of these multisymplectic manifolds characterizes
%their multisymplectic diffeomorphisms \cite{EIMR-if}.
%
Nevertheless, further developments have not been achieved.
For instance, concerning reduction theory, only partial results
about reduction by foliations are currently being studied \cite{Ib-2000}.
The theory of reduction of systems of ``multisymplectic'' {\sc pde}s under the action of
groups of symmetries, obtaining other simpler but
also ``multisymplectic'' ones, is under research 
\cite{CRS}, \cite{Hr-99}, \cite{MW}.

In the same way, approaches for generalizing symplectic integrators
to this geometric framework (i.e., the so-called {\sl multisymplectic integrators\/})
have only been recently developed \cite{MPS-mi}.

Other results similar to those stated in Theorems \ref{KKS}
(Konstant-Kirillov-Souriau) and \ref{AL}
(Arnold-Liouville) have not been achieved yet. 

One can expect to see more work on all these subjects in the future.

\subsection*{Acknowledgments}

We acknowledge the financial support of {\sl Ministerio de Ciencia y Tecnolog\'\i a},
 BFM2002-03493.
We wish to thank Mr. Jeff Palmer for his
assistance in preparing the English version of the manuscript.

%\newpage

%%%%%%%%%%%%%%%%%%%%%%%%%%%%%%%%%%%%%%%%%%%%%%%%%%%%%%%%%%%%%%%%%%%%%%%%%%%%%%%%%%%%%%%%%%%%%%%%%%%%%
\begin{thebiblio}{03}
\bibitem{AM}
R. Abraham, J.E. Marsden,
{\it Foundations of Mechanics\/} (2nd ed.),
Addison-Wesley, Reading Ma., 1978.
\bibitem{Ar} 
V.I. Arnold, {\it Mathematical methods of classical mechanics}. 
Graduate Texts in Mathematics {\bf 60}. Springer-Verlag, New York, 1989.
\bibitem{Aw-92}
A. Awane, ``$k$-symplectic structures'', {\it J. Math. Phys.} {\bf 32}(12) (1992) 4046-4052.
\bibitem{Ba97}
A. Banyaga, {\it The structure of classical
diffeomorphism groups}. Mathematics and its Applications {\bf 400}.
 Kluwe Acad. Pub. Group., Dordrecht (1997) 113-118.
\bibitem{BS-01}
{\rm G. Blankenstein and A.\,J. van der Schaft},
``Symmetry and reduction in implicit generalized Hamiltonian systems'',
{\it Rep.\ Math.\ Phys.~\bf 47} (2001) 57-100. 
\bibitem{BS}
A.I. Bobenko, Y.B. Suris,
``Discrete Lagrangian reduction, discrete Euler-Poincar\'e equations and
semidirect products'', {\it Lett. Math. Phys}. {\bf 49} (1999) 79-93.
\bibitem{Ca-80}
{\rm S.\,L. Campbell},
{\it Singular systems of differential equations},
Research Notes in Mathematics,
Pitman, San Francisco, 1980.
\bibitem{CHYZ}
{\rm S.\,L. Campbell, R. Hollenbeck, K. Yeomans, Y. Zhong},
``Mixed symbolic-numerical computations with general {\sc dae}s I:
System properties'',
{\it Numerical Algorithms} {\bf 19} (1998) 73-83.
\bibitem{Can} 
A. Cannas da Silva, {\it Lectures on Symplectic Geometry}
Springer-Verlag, Berlin, Heidelberg, 2001.
\bibitem{CIL-96}
F. Cantrijn, L.A. Ibort, M. de Le\'on, ``Hamiltonian
Structures on Multisymplectic Manifolds'', {\it Rnd. Sem. Math.
Univ. Pol. Torino} {\bf 54}, (1996) 225-236.
\bibitem{CIL-98}
F. Cantrijn, L.A. Ibort, M. de Le\'on, ``On the Geometry of
Multisymplectic Manifolds'', {\it J. Austral. Math. Soc. Ser.}
{\bf 66} (1999) 303-330.
\bibitem{Ca-sl}
J.F. Cari\~nena,
``Theory of singular Lagrangians'', 
{\it Fortschr. Phys.} {\bf 38}(9) (1990) 641--679.
\bibitem{CCI-91}
J.F. Cari\~nena, M. Crampin,  L.A. Ibort,
``On the multisymplectic formalism for first order field theories'',
{\it Diff. Geom. Appl.} {\bf 1} (1991) 345-374.
\bibitem{CGIR-85}
J.F. Cari\~nena, J. Gomis, L.A. Ibort, N. Rom\'an-Roy,
``Canonical transformation theory for presymplectic systems'',
{\it J. Math. Phys.} {\bf 26} (1985) 1961-1969.
\bibitem{CRS}
M. Castrill\'on-L\'opez, T.S. Ratiu, S. Shkoller,
``Reduction in principal fiber bundles: Covariant Euler-Poincr\'e
equations'', Amer. Math. Soc., Providence, RI.,
{\it Proc. Amer. Math. Soc.} (2000) 2155-2164.
\bibitem{CO-88}
{\rm L.\,O. Chua, H. Oka},
``Normal forms for constrained nonlinear differential equations.
Part~I: theory'',
{\it IEEE Trans. Circuits Syst.~\bf 35} (1988) 881-901.
\bibitem{CM}
J. Cort\'es, S. Mart\'\i nez, 
``Non-holonomic integrators'', {\it Nonlinearity} {\bf 14}(5) (2001) 1365-1392.
\bibitem{CB}
R. Cushman, L. Bates,
{\it Global Aspects of Classical Integrable Systems}.
Birkhauser, Boston 1976.
\bibitem{Dir-lectures} 
{\rm P.\,A.\,M. Dirac}, 
{\it Lectures on Quantum Mechanics}, 
Yeshiva University, New York, 1964.
%\bibitem{EIMR-if}
%A. Echeverr\'\i a-Enr\'\i quez, L.A. Ibort, M.C. Mu\~noz-Lecanda, N. Rom\'an-Roy.
%``Invariant Forms and Automorphisms of a class of Multisymplectic Manifolds''.
%math-ph/9805040 (1998).
\bibitem{EMR-96}
A. Echeverr\'\i a-Enr\'\i quez, M.C. Mu\~noz-Lecanda, N. Rom\'an-Roy,
``Geometry of Lagrangian first-order classical field theories''.
{\it Forts. Phys.} {\bf 44} (1996) 235-280.
\bibitem{EMR-98}
A. Echeverr\'\i a-Enr\'\i quez, M.C. Mu\~noz-Lecanda, N. Rom\'an-Roy,
``Multivector Fields and Connections.
Setting Lagrangian Equations in Field Theories''.
{\it J. Math. Phys.} {\bf 39}(9) (1998) 4578-4603.
\bibitem{EMR-99}
A. Echeverr\'\i a-Enr\'\i quez, M.C. Mu\~noz-Lecanda, N.
Rom\'an-Roy, ``Reduction of presymplectic manifolds with
symmetry'', {\it Rev. Math. Phys.} {\bf 11}(10) (1999) 1209-1247.
\bibitem{EMR-00}
A. Echeverr\'\i a-Enr\'\i quez, M.C. Mu\~noz-Lecanda, N.
 Rom\'an-Roy, ``Geometry of Multisymplectic Hamiltonian First-order
 Field Theories'', {\it J. Math. Phys.} {\bf 41}(11) (2000) 7402-7444.
\bibitem{EMR-gq} 
A. Echeverr\'\i a-Enr\'\i quez, M.C. Mu\~noz-Lecanda,
N. Rom\'an-Roy, C. Victoria, ``Mathematical foundations of geometric quantization'', 
{\it Extracta Mathematicae}, {\bf 13}(2), 135-238, (1998).
\bibitem{Gc-73}
P.L. Garc\'\i a, ``The Poincar\'e-Cartan invariant in the
calculus of variations'', {\it Symp. Math.} {\bf 14} (Convegno di
Geometria Simplettica e Fisica Matematica, INDAM, Rome, 1973),
Acad. Press, London  (1974) 219-246.
\bibitem{GMS-97}
G. Giachetta, L. Mangiarotti, G. Sardanashvily, {\it New
Lagrangian and Hamiltonian Methods in Field Theory}, World
Scientific Pub. Co., Singapore (1997).
\bibitem{GS-73}
H. Goldschmidt, S. Sternberg, ``The Hamilton-Cartan
formalism in the calculus of variations'', {\it Ann. Inst. Fourier
Grenoble} {\bf 23}(1) (1973) 203-267.
\bibitem{GLR-84}
J. Gomis, J. Llosa, N. Rom\'an-Roy, 
``Lee Hwa Chung theorem for presymplectic manifolds.  Canonical
transformations for constrained systems'', {\it J. Math. Phys.} {\bf 25} (1984) 1348-1355.
\bibitem{Go-82}
M.J. Gotay,
``On coisotropic imbeddings of presymplectic manifolds'',
{\it Proc. Amer. Math. Soc.} {\bf 84} (1982) 111-114.
\bibitem{GN-80a}
M.J. Gotay, J.M. Nester, ``Presymplectic Lagrangian systems
II: the second order equation problem'', {\it Ann. Inst. H.
Poincar\'e A} {\bf 32} (1980) 1-13.
\bibitem{GN-80} 
{\rm M.\,J. Gotay, J.\,M. Nester}, 
``Presymplecticlagrangian systems II: the second order equation
problem'', 
{\it Ann. Inst. H.~Poincar\'e~A~\bf32} (1980) 1--13.
\bibitem{GNH-78}
{\rm M.\,J.~Gotay, J.\,M.~Nester, G.~Hinds},
``Presymplectic manifolds and the Dirac--Bergmann theory of constraints'',
{\it J.~Math. Phys.~\bf 27} (1978) 2388-2399.
\bibitem{GMR-96} 
{\rm X. Gr\`acia, M.C. Mu\~noz-Lecanda, N. Rom\'an-Roy}, 
``Singular systems: their origin, gneral features and non-numerical solution'', 
{\it Proc. of the IFAC Conference on System, Structure and Control},
Nantes, France 1995. M. Guglielmi ed., Pergamon (1996) 31-36.
\bibitem{GP-92} 
{\rm X. Gr\`acia, J.\,M. Pons}, 
``A generalized geometric framework for constrained systems'',
{\it Diff. Geom. Appl.~\bf 2} (1992) 223-247.
\bibitem{GP-01} 
{\rm X. Gr\`acia, J.\,M. Pons}, 
``Singular lagrangians: some geometric structures along the Legendre map'', 
{\it J.~Phys.~A: Math. Gen.~\bf 34} (2001) 3047-3070.
\bibitem{GP-02} 
{\rm X. Gr\`acia, J.\,M. Pons}, 
``Symmetries and infinitesimal symmetries of singular differential 
equations'', 
{\it J.~Phys.~A: Math. Gen.~\bf 35} (2002) 5059-5077.
\bibitem{GPR-91}
{\rm X. Gr\`acia, J.\,M. Pons, N. Rom\'an-Roy}, 
``Higher order lagrangian systems:
geometric structures, dynamics and constraints'',
{\it J.~Math. Phys.~\bf 32} (1991) 2744-2763.
\bibitem{Gu-87}
C G\"unther, ``The polysymplectic Hamiltonian formalism in
the Field Theory and the calculus of variations I: the local
case'', {\it J. Diff. Geom.} {\bf 25} (1987) 23-53.
\bibitem{HB-84} 
{\rm B.\,C. Haggman, P.\,P. Bryant}, 
``Solutions of singular constrained differential equations:
a generalization of circuits containing capacitor-only loops and 
inductor-only cutsets'',
{\it IEEE Trans. Circuits Syst. \bf CAS-31}(12) (1984) 1015-1025.
\bibitem{HLR-89}
{\rm E. Hairer, C. Lubich, M. Roche}, 
{\it The Numerical Solution of Differential-Algebraic
Systems by Runge-Kutta Methods}, 
Lecture Notes inMathematics {\bf 1409}, Springer, Berlin, 1989.
\bibitem{HW-91} 
{\rm E. Hairer, G. Wanner}, 
{\it Solving Ordinary Differential Equations~II}, 
Springer, Berlin, 1991.
\bibitem{HK-02}
F. H\'elein, J. Kouneiher,
``Finite dimensional Hamiltonian formalism for gauge and quantum field theories'',
{\it  J. Math. Phys.} {\bf 43}(5) (2002) 2306--2347.
\bibitem{Hr-99}
S.P. Hrabak. ``On a Multisymplectic Formulation of
the Classical BRST Symmetry''. Ph. D Thesis,
Dpt. Mathematics, King's College, Strand (1999).
\bibitem{Hw-47} 
L. Hwa Chung, ``The Universal Integral Invariants of
Hamiltonian Systems and Applications to the Theory of Canonical
Transformations'', {\it Proc. Roy. Soc.} {\bf LXIIA} (1947) 237-246.
\bibitem{Ib-2000}
L.A. Ibort,
``Multisymplectic manifolds: general aspects and particular situations'',
{\it Proc. IX Fall Workshop on Geometry and Physics},
Vilanova i la Geltr\'u, Spain (2000).
Pubs. RSME {\bf 3}; X. Gr\`acia, J. Mar\'\i n-Solano,
M.C. Mu\~noz-Lecanda, N. Rom\'an-Roy Eds.
Madrid (2001) 79-88.
\bibitem{Isi}
{\rm A. Isidori},
{\it Nonlinear control systems}, 3rd ed.,
Springer, Berlin, 1995.
\bibitem{Kac} 
{\rm T. Kaczorek}, 
{\it Linear control systems}, vol.~I, 
Research Studies Press, Taunton, 1992. 
\bibitem{Ka-98}
I.V. Kanatchikov, ``Canonical structure of Classical Field
Theory in the polymomentum phase space'', {\it Rep. Math. Phys.}
{\bf 41}(1) (1998) 49-90.
\bibitem{KT-79}
{\rm J. Kijowski, W.M. Tulczyjew}, {\it A Symplectic Framework for
Field Theories}, Lect. Notes Phys. {\bf 170}, Springer-Verlag,
Berlin 1979.
\bibitem{Ki-gq}
A.A. Kirillov,
``Geometric Quantization'',
in the {\it Encyclopaedia of Mathematical Sciences\/}
(vol 4): {\it Dynamical Systems\/},
Springer, Berlin (1985) 137-172.
\bibitem{Kos}
B. Kostant,
``Orbits, symplectic structures and representation theory''.
{\it Proc. US-Japan Seminar on Diff. Geom.} (Kyoto),
Nippon Hyronsha, Tokyo {\bf 77}, 1966.
\bibitem{Ko-70}
B. Kostant,
``Quantization and unitary representations'',
{\it Lects. in Mod. Anal. and Appl. III},
Lecture Notes in Mathematics {\bf 170}, Springer, New York (1970)
87-208.
\bibitem{Kr-02}
O. Krupkova, ``Hamiltonian field theory'',
{\it J. Geom. Phys.} {\bf 43}(2-3) (2002) 93-132.
\bibitem{LMM-96}
M. de Le\'on, J. Mar\'\i n-Solano, J.C. Marrero,
``A Geometrical approach to Classical Field Theories: A constraint
algorithm for singular theories'', 
Proc. on {\it New Developments in Differential Geometry},
L. Tamassi-J. Szenthe eds., Kluwer Acad. Press, (1996) 291-312.
\bibitem{LMMMR}
{\rm M. de Le\'on, J. Mar\'\i n-Solano, J.C. Marrero, M.C. Mu\~noz-Lecanda,
N. Rom\'an-Roy},
``Singular Lagrangian Systems on Jet bundles'',
{\it Fortschr. Phys.} {\bf 50}(2) (2002) 105-169.
\bibitem{LMd}
M. de Le\'on, D. Mart\'\i n de Diego,
`` Variational integrators and time-dependent Lagrangian systems''.
XXXIII Symposium on Mathematical Physics (Tor\'un, 2001). 
{\it Rep. Math. Phys.} {\bf 49}(2-3) (2002) 183-192. 
\bibitem{LMS-2003}
M. de Le\'on D. Mart\'\i n de Diego, A. Santamar\'\i a-Merino,
``Tulczyjew triples and Lagrangian submanifolds in classical field theories'',
in {\it Applied Differential Geometry and Mechanics};
W. Sarlet and F. Cantrijn eds. Univ. of Gent, Gent, Academia Press (2003)
21-47.
\bibitem{LMS-99}
M. de Le\'on, E. Merino, M. Salgado,
 ``$k$-cosymplectic Manifolds and Lagrangian
 Formalism for Field Theories''. {\it J. Math. Phys.} {\bf 42}(5) (2001) 2092-2104.
\bibitem{LM-87}
{\rm P. Libermann, C.M. Marle},
{\it Symplectic geometry and analytical dynamics\/},
D. Reidel Publisging Company, Dordrecht, 1987.
\bibitem{Lic-75} 
{\rm A. Lichnerowicz}, 
``Vari\'et\'e symplectique et dynamique associ\'ee \`a une
sous-vari\'et\'e'', 
{\it C.\,R.~Acad. Sc. Paris~\bf 280}(A) (1975) 523--527.
%\bibitem{LlR-88} 
%J. Llosa, N. Rom\'an-Roy, ``Invariant forms and Hamiltonian systems: 
%A Geometrical Setting'', 
%{\it Int. J. Theor. Phys.} {\bf 27}(12) (1988) 1533-1543.
\bibitem{MS-98} 
L. Mangiarotti, G. Sardanashvily, 
{\it Gauge Mechanics}. 
World Scientific, Singapore, 1998.
\bibitem{Ma-83}
C.L. Marle,
``Sous-vari\'et\'es de rang constant d'une variet\'e symplectique'',
{\it Ast\'erisque} {\bf 107-108} (1983) 69-87.
\bibitem{MMT-92} 
{\rm G. Marmo, G. Mendella, W.\,M. Tulczijew},
``Symmetries and constant of the motion for dynamics in implicit form'', 
{\it Ann. Inst. H.~Poincar\'e~A~\bf 57} (1992) 147-166.
\bibitem{MMT-97} 
{\rm G. Marmo, G. Mendella and W.\,M. Tulczijew},
``Constrained hamiltonian systems as implicit differential equations'', 
{\it J.~Phys.~A: Math. Gen. \bf 30} (1997) 277--293.
\bibitem{Ma-92} 
J.E. Marsden, {\it Lectures on Mechanics},
London Math. Soc. Lecture Note Ser. {\bf 174}, Cambridge Univ. Press,
 Cambridge, 1992.
\bibitem{MPS-mi}
J.E. Marsden, S. G.W. Patrick, S. Shkoller, 
``Multisymplectic geometry, variational integrators and nonlinear {\sc pde}s'',
{\it Comm. Math. Phys.} {\bf 199} (1998) 351-395.
\bibitem{MPS}
J.E. Marsden, S. Pekarsky, S. Shkoller, 
``Discrete Euler-Poincar\'e and Lie-Poisson equations'', {\it Nonlinearity}
{\bf 12} (1999) 1647-1662.
\bibitem{MS-99}
J.E. Marsden, S. Shkoller, ``Multisymplectic Geometry,
Covariant Hamiltonians and Water Waves'', {\it Math. Proc. Camb.
Phil. Soc.} {\bf 125} (1999) 553-575.
\bibitem{MW-74} 
J.E. Marsden, A. Weinstein, ``Reduction of symplectic manifolds with symmetry'',
{\it Rep. Math. Phys.} {\bf 5} (1974) 121-130.
\bibitem{MW} 
J.E. Marsden, A. Weinstein, ``Some comments on the History,
Theory, and Applications of Symplectic Reduction'',
{\it Quantization of singular symplectic quotiens}.
N. Landsman, M. Pflaum, M. Schlichenmanier eds., Birkhauser, Boston
(2001) 1-20.
\bibitem{Ma70} 
J. Martinet,
``Sur les singularit\'es des formes diff\'erentielles'', 
{\it Ann. Inst. Fourier, Grenoble} {\bf 20} (1) (1970) 95-178.
\bibitem{Sa} 
D. McDuff, D. Salamon, {\it Introduction to symplectic topology},
Oxford Clarendon Press, 1998.
\bibitem{Med} 
{\rm M. Medved'}, 
``Normal forms of implicit and observed implicit differential equations'', 
{\it Riv. Mat. Pura Appl. \bf 10} (1990), 95-107
\bibitem{MMT-95} 
{\rm G. Mendella, G. Marmo and W.\,M. Tulczijew},
``Integrability of implicit differential equations'', 
{\it J.~Phys.~A: Math. Gen. \bf 28} (1995) 149-163.
\bibitem{MT-78}
{\rm M.\,R. Menzio, W.\,M. Tulczijew},
``Infinitesimal symplectic relations and generalized hamiltonian 
dynamics'',
{\it Ann. Inst. H.~Poincar\'e~A~\bf 28} (1978) 349--367.
\bibitem{MR-92}
{\rm M.\,C. Mu\~noz-Lecanda, N. Rom\'an-Roy},
``Lagrangian theory for presymplectic systems'', 
{\it Ann. Inst. H. Poincar\'e} {\bf 57} (1992) 27-45. 
\bibitem{MR-99}
{\rm M.\,C. Mu\~noz-Lecanda, N. Rom\'an-Roy},
``Implicit quasilinear differential systems: a geometrical approach'', 
{\it Electron. J. Differential Equations \bf 1999}, No.~10, 33~pp. 
\bibitem{No-93}
L.K. Norris, ``Generalized Symplectic Geometry in the Frame Bundle of a Manifold'',
 {\it Proc. Symposia in Pure Math.} {\bf 54}(2) (1993) 435-465.
\bibitem{PR-02}
C. Paufler, H. Romer, 
``Geometry of Hamiltonian $n$-vector fields in multisymplectic field theory'',
{\it J. Geom. Phys.} {\bf 44}(1) (2002) 52-69.
\bibitem{RR-91}
{\rm P.\,J. Rabier, W.\,C. Rheinboldt},
``A general existence and uniqueness theory for implicit
differential-algebraic equations'',
{\it Differential Integral Equations~\bf 4} (1991) 563-582.
\bibitem{RR-94}
{\rm P.\,J. Rabier, W.\,C. Rheinboldt}, 
``A geometric treatment of implicit differential-algebraic equations'',
{\it J.~Differential Equations~\bf 109} (1994) 110-146.
\bibitem{Rei} 
{\rm S. Reich}, 
``On a geometrical interpretation of differential-algebraic equations'', 
{\it Circuits Systems Signal Process. \bf 9} (1990) 367-382. 
 \bibitem{Rhe-84}
{\rm W.\,C. Rheinboldt},
``Differential-algebraic systems as differential equations on manifolds'',
{\it Math. Comp.~\bf 43}(168) (1984) 473-482.
\bibitem{RZ} 
{\rm R. Riaza, P.\,J. Zufiria}, 
``Stability of singular equilibria in quasilinear implicit differential 
equations'', 
{\it J.~Differential Equations~\bf 171} (2001) 24-53.
\bibitem{Sd-95}
G. Sardanashvily,
{\it Generalized Hamiltonian Formalism for Field Theory. Constraint Systems},
World Scientific, Singapore (1995).
\bibitem{Sa-89}
D.J. Saunders,
{\it The Geometry of Jet Bundles},
London Math. Soc. Lect. Notes Ser.
{\bf 142}, Cambridge, Univ. Press, 1989.
\bibitem{SR-83} 
{\rm R. Skinner, R. Rusk}, 
``Generalized hamiltonian dynamics~I: formulation on $T^*Q\otimes TQ$'', 
{\it J.~Math. Phys.~\bf 24} (1983) 2589-2594.
\bibitem{Sni-74} 
{\rm J.\'Sniatycki}, 
``Dirac brackets in geometric dynamics'', 
{\it Ann. Inst. H.~Poincar\'e~A~\bf 20} (1974) 365-372.
\bibitem{Sn-80}
J. \'Sniatycki,
{\it Geometric quantization and quantum mechanics\/},
Springer-Verlag, Berlin, 1980.
\bibitem{So-66}
J.M. Souriau,
``Quantification g\'eom\`etrique'',
{\it Comm. Math. Phys.} {\bf 1}(11) (1966) 374-398.
Dunod, Paris, 1970.
\bibitem{Sou}
J.M. Souriau,
{\it Structure des Syst\`emes Dynamiques}.
Dunod, Paris, 1970.
\bibitem{Tak-76}
{\rm F. Takens},
``Constrained equations: a study of implicit differential equations 
and their discontinuous solutions'', in 
{\it Structural Stability, the Theory of Catastrophes and
Application in the Sciences}, P. Hilton ed.,
Lectures Notes in Mathematics {\bf 525},
Springer-Verlag, Berlin, 1976.
\bibitem{Tu-76a}
W.M. Tulczyjew,
``Les sous-vari\'et\'es lagrangiennes et la dynamique hamiltonniene'',
{\it C.R. Acad. Sci. Paris Ser. A} {\bf t 283}(1) (1976) 15-18.
\bibitem{Tu-76b}
W.M. Tulczyjew,
``Les sous-vari\'et\'es lagrangiennes et la dynamique lagrangiene'',
{\it C.R. Acad. Sci. Paris Ser. A} {\bf t 283}(8) (1976) 675-678.
\bibitem{Tu-77}
W.M. Tulczyjew,
``The Legendre transformation''. 
{\it Ann. Inst. H. Poinca\'e Sect. A} {\bf 27}(1) (1977) 101-114. 
\bibitem{Tuy} 
G. Tuynman, ``What is prequantization, and what is geometric quantization?'',
{\it Proceedings Seminar 1989-1990 Mathematical Structures in Field Theory (Amsterdam)},
(1990) 1-28.
\bibitem{VK}
A.M. Vinogradov, B.A. Kupershmidt,
``The structure of Hamiltonian mechanics'',
{\it Uspehi Mat. Nauk} {\bf 32}(4) (1977) 175-236.
\bibitem{We-2}
A. Weinstein,
``Symplectic Manifolds and their Lagrangian submanifolds''.
{\it Adv. in Math.} {\bf 6} (1971) 329-346.
\bibitem{We}
A. Weinstein,
{\it Lectures on Symplectic Manifolds}.
CBMS Conf Ser., Am. Math. Soc. {\bf 29}. Am. Math. Soc., Providence, RI, 1977.
\bibitem{Wh} 
E.T. Whittaker, 
{\it A treatise on the analytical dynamics of particles and rigid bodies}. 
Cambridge Univ. Press (4th Ed.), New York, 1988.
\bibitem{Wi1}
J. Williamson,
``On the algebraic problem concerning the normal forms of linear dynamical systems'',
{\it Am. J. Math.} {\bf 58} (1936) 141-163.
\bibitem{Wo} 
N.M. Woodhouse, {\it Geometric Quantization}, Clarendon Press, Oxford,
1992. 
\end{thebiblio}
%%%%%%%%%%%%%%%%%%%%%%%%%%%%%%%%%%%%%%%%%%%%%%%%%%%%%%%%%%%%%%%%%%%%%%%%%%%%%%%%%%%%%%%%%%%%%%%%%%%%%

\end{document}